\documentclass[aps,prb,twocolumn,superscriptaddress,showpacs]{revtex4}
\usepackage{ae}
\usepackage[T1]{fontenc}
\usepackage[ansinew]{inputenc}
\usepackage{amsmath}
\usepackage{amssymb}
\usepackage{graphicx}
\usepackage{color}
\usepackage[colorlinks]{hyperref}
\usepackage{epstopdf}

\begin{document}

\date{\today}

\title{Density wave and topological reconstruction of an isotropic two-dimensional electron band in external magnetic field}

\author{A.M. Kadigrobov* }
\affiliation{Theoretische Physik III, Ruhr-Universitaet Bochum, D-44801 Bochum, Germany}
\affiliation{Department of Physics, Faculty of Science, University of Zagreb, Croatia}
\author{D. Radi\'{c}*}
\affiliation{Department of Physics, Faculty of Science, University of Zagreb, Croatia}
\author{A. Bjeli\v{s}}
\affiliation{Department of Physics, Faculty of Science, University of Zagreb, Croatia}

%\date{\today}

\begin{abstract}

We predict a mechanism of spontaneous stabilization of a uniaxial density wave in a
two-dimensional metal with an isotropic Fermi surface in the presence of external
magnetic field. The topological transformation of a closed Fermi
surface into an open one decreases the electron band energy due to delocalization
of electrons initially localized by magnetic field, additionally affected by the
magnetic breakdown effect. The driving mechanism of such reconstruction is
a periodic potential due to the self-consistently formed electron density wave. It
is accompanied with quantum oscillations periodic in inverse magnetic field, similar
to the standard de Haas - van Alphen effect, due to Landau level filling. The phase
transition appears as a quantum one at T=0, provided the relevant coupling
constant is above the critical one. This critical value rapidly decreases,
and finally saturates toward zero on the scale of tens of Tesla. Thus, a strong enough
magnetic field can induce the density wave in the system in which it was absent in zero
field.

\end{abstract}

\maketitle

\section{Introduction}

The instability of low dimensional conductors with the spontaneous arising of a periodic
modulation of the crystal, usually called the density wave (DW) \cite{Gruener}, remains in the
focus of attention since its early prediction by Peierls\cite{Peierls} almost ninety years ago.
In one-dimensional (1D) conductors the crystal modulation opens a gap in the electron band at
the initial Fermi energy, decreasing so the electron band energy. The new DW ordering is stabilized
whenever this energy decrease overwhelms the competing increase of the crystal energy caused by the
accompanying lattice modulation.

The DW instability also arises in the special class of two-dimensional (2D) systems, often also called
quasi-one-dimensional (quasi-1D) systems, with a highly anisotropic, mainly open Fermi surface (FS),
such that the parts of its contour can be (almost) perfectly mapped, i. e. nested,
onto each other. Density waves of this type have been intensively investigated and widely observed in
the series of different materials possessing such band dispersions \cite{Gruener,Pouget}.

However, DWs have not been observed only in the conductors with highly anisotropic Fermi surfaces,
but also in many 2D conductors with closed Fermi surfaces for which the nesting condition as
specified above is far from being fulfilled. Particularly significant in this respect are
high-temperature superconducting  cuprates  with conducting $CuO_{2}$ layers \cite{cuprates}, as
well as hexagonal (semi)metallic layers appearing in graphene-based intercalates, like in e. g.
$CaC_{6}$ \cite{graphene}. Despite intensive investigations of these and similar materials, the
origin of the observed structural instability is still a controversial topic.

\begin{figure}
\centerline{\includegraphics[width=\columnwidth]{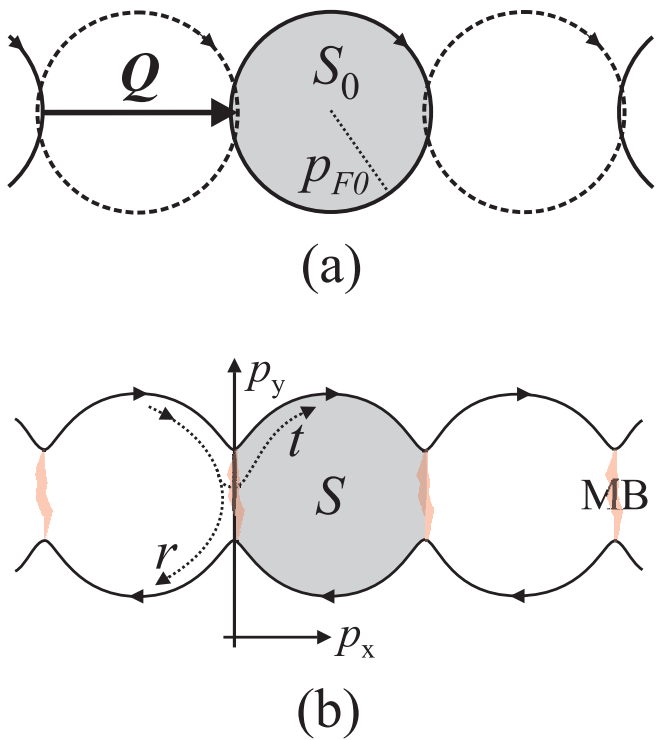}}
\caption{(a) The reciprocal space for the isotropic 2D band with the Fermi momentum $p_{F0}$ and the DW which
introduces the 1D reciprocal lattice with the wave number $Q/\hbar$ close to $2p_{F0}/\hbar$.
(b) Two open lines representing the reconstructed FS due to the finite amplitude of the DW in the presence
of finite magnetic field directed perpendicularly to the plane $(p_{x}, p_{y})$, enabling
finite MB probability amplitudes $t$ and $r$.}
\label{trajectories}
\end{figure}

In our recent paper \cite{KBR} we have proposed a mechanism of the DW ordering
based on the topological reconstruction of the highly symmetric, initially closed Fermi surface. As shown
in Fig.\ref{trajectories}(a), the DW has a wave vector $\mathbf{Q}/\hbar$ that brings the initial FS
into the so-called touching range generated at the edge of the Brillouin zone
established by the DW periodic
modulation in the system. By lifting the energy degeneracy and opening the gap in that region, the
band topology changes: two parabolic initial bands now form the lower band with saddle point and
the upper band with parabolic minimum at the touching point\cite{KBR}. The new Fermi
surface then becomes open as shown in Fig.\ref{trajectories}(b), while the total band
energy is decreased. In principle this decrease may stem from
two contributions: lowering of the new Fermi energy with respect to the original one, and the
change in the density of states due to the redistribution of filled states from higher towards
lower energies. It appears that in the particular analyzed case\cite{KBR}, and
for the optimal DW wave vector, there is only the contribution that
comes from the redistribution of states, while the Fermi energy shows no change.

Let us now focus on the role that the external magnetic field may have in the DW ordering
and the accompanying band reconstruction. Namely, as it has been observed in various 2D
conductors, in particular in high-temperature cuprates \cite{LeBoeuf,Wu,Laliberte}, it appears
that DWs also spontaneously arise under a strong enough external magnetic field. These magnetic
field-induced DWs have been most often associated to the Landau quantization of orbits in relatively
small pockets in the anisotropic bands \cite{Gorkov-Lebed, HML,RBZ1}. Within this framework,
particularly interesting is the role of magnetic breakdown due to the tunneling of electrons through
the narrow barriers in the reciprocal space, quite often present in the band spectra
under question \cite{Gorkov-Lebed-mb,RBZ2}. In fact, in few earlier papers \cite{PRL,EPJB} we pointed
out that the electron delocalization due to such tunneling could itself have the decisive role in
the stabilization of the so-called magnetic breakdown induced density waves (MBIDWs).

%%%%%%%%%%%%%%%%%%%%%%%%%%
%%%%%%%%%%%%%%%%%%%%%%%%%%%%

Our prediction was based on the analysis of 2D conductors with open Fermi surfaces (as is the case
of e. g. some compounds from the family of Bechgaard salts) in the presence of strong external magnetic
fields \cite{Lebed}. The physical mechanism causing this phase transition is specific with respect to
those of the standard Peierls transition in quasi-one-dimensional conductors. Namely, under a strong
magnetic field electrons of the 2D conductor with the open Fermi surface move
in the opposite directions along two trajectories defined by the equation
$\varepsilon(\textbf{\emph{p}})= \varepsilon_{F}$, where $\varepsilon(\textbf{\emph{p}})$ defines the
band dispersion, $\varepsilon_{F}$ is Fermi energy, and space inversion is assumed. As it was suggested
in the papers \cite{PRL,EPJB}, the periodic modulation of electronic charge caused by the DW transforms the open
Fermi surface into a periodic chain of overlapping above-mentioned trajectories with lifted band degeneracy
in the crossing points of two initially open sub-bands. As a result, after the band reconstruction, electrons
move in the presence of magnetic field along one-dimensional chain of trajectories with periodically
located scatterers - small areas around the crossing points at which the magnetic breakdown
\cite{Falicov,Kaganov}, the quantum tunneling between the neighboring trajectories, takes place. In other
words, within the reconstructed Fermi surface electrons under magnetic field  move along a one-dimensional
periodic set of quantum barriers and hence the system is mapped onto the one-dimensional metal with the
"crystal" period of the order of the Larmor radius $R_H=c \overline{p}_F/eH$, where $H$ is the magnetic
field, $\overline{p}_F$ is the value of the longitudinal Fermi wave vector averaged over the reciprocal space,
$e$ is the electron charge, and $c$ is the velocity of light. The emerging electron spectrum will thus be a
set of alternating narrow energy bands and energy gaps. The transition of Peierls type in such a system takes
place provided the initial Fermi energy is inside one of the gaps in the new spectrum.

Evidently, since the above physical considerations and results of the cited papers have the
quasi-two-dimensional band dispersions with open Fermi surfaces as the starting point, they can not be
straightforwardly transposed to highly isotropic 2D conductors with closed Fermi surfaces. Still, as
stated above, the band reconstruction due to a finite DW introduces here as well the one-dimensional
periodic set of barriers, and the quantum tunneling in
the finite magnetic field is again unavoidable. In the present paper we consider the latter situation.
Extending the treatment initiated in our previous work \cite{KBR}, we show that in such situation one can
stabilize the DW under a strong magnetic field as well. Even more, we show that under magnetic field the
DW order is additionally strengthened, and appears at lower values of the critical coupling constant.

The underlying physical reason for this stabilization is as follows. As already mentioned, the DW causes
a topological band reconstruction under which the initially closed two-dimensional FS
(Fig.\ref{trajectories}(a)) is transformed into an open one (Fig.\ref{trajectories}(b)). Let us now impose
the perpendicular magnetic field and consider its effects within the semi-classical reasoning. Without the
band reconstruction, electrons would move along initially closed Fermi surface, with the discrete spectrum
consisting of degenerate Landau levels, and with an increase of the band energy in average. However, having
the reconstruction, the semi-classical motion along open trajectories delocalizes band states and decreases
their energy. This delocalization itself would therefore act towards an additional stabilization of the DW, beside
that realized due to the band reconstruction in the absence of magnetic field.

However, the picture of electron dynamics is not completed by this. As it is seen in
Fig.\ref{trajectories}(b), two open trajectories are close to each other in the touching range of the
reconstructed band, enabling in this range the magnetic breakdown (MB), i. e. the tunneling of electrons from
one open trajectory to another. The finite tunneling is characterized by two probability amplitudes for an electron,
$t$ to pass through the barrier and continue the motion along the same open trajectory, and $r$ to
get reflected back and start to move in the opposite direction along the second trajectory, with $t^2+r^2=1$. The MB due to
the finiteness of $r$ partially re-localizes the effective electron states and increases the corresponding
band energy, thus acting against the DW stabilization.

Aiming to make the quantitative analysis of these two opposite tendencies, we formulate
the appropriate mean-field approach and undertake
the detailed calculation of the band spectrum and of the corresponding total energy of ground state in the
presence of external magnetic field. By this we arrive to the conditions for the stabilization of the DW
and reconstructed band in the magnetic field. In particular, we determine the range of the effective strength
of electron-phonon coupling (or of the coupling to some other effective or real boson field), under which the
DW ordering takes place in the given magnetic field.

In Sec. II we start with the electron-phonon Hamiltonian in the presence of
finite magnetic field, and introduce the mean-field approximation relevant for the stabilization
of DW and band reconstruction. The impact of the magnetic field on the band reconstruction, which
includes the semiclassical treatment of uniaxial electron trajectories and the magnetic breakdown in
the ranges of topological change of initial 2D band, is treated in Sec. III. In Sec. IV we calculate
the total band energy, while in Sec. V the minimization of the total DW condensation energy leads us
to the determination of the optimal amplitude and the wave number of the DW order, and their dependence on
the magnetic field and the electron-phonon coupling constant. Sec. VI contains concluding remarks.

\section{Hamiltonian and mean-field approximation}

For the sake of simplicity we consider a 2D-conductor with the electron band which initially has a parabolic
dispersion $\epsilon_0(k_x,k_y) =(k_x^2+k_y^2)/2m$, where $\mathbf{k}=(k_x,k_y)$ is the electron momentum and
$m$ is the effective electron mass. Taking the Hamiltonian describing coupled electron-phonon system in the
momentum representation, and making the Peierls substitution with the momentum operator $\hat{\mathbf{k}}$
replaced by the operator  $\hat{\mathbf{k}}-(e/c)\mathbf{ A}$ where $\mathbf{ A}$ is the vector potential,
one gets
\begin{eqnarray}
\mathcal{H}=\frac{1}{2 m}\sum_{\mathbf{k}}a^{\dag}_\mathbf{k}\left[ k_x^2+\left(k_y+i \frac{e
\hbar
H}{c}\frac{\partial}{\partial k_x}\right)^2  \right]a_\mathbf{k} \nonumber \\
+ \sum_{\mathbf{q}}\hbar \omega(\mathbf{q})b^{\dag}_\mathbf{q}b_\mathbf{q} +
\frac{1}{\sqrt{A}}g\sum_{\mathbf{k},\mathbf{q}}
a^{\dag}_{\mathbf{k}+\mathbf{q}}a_\mathbf{k}\left(b^{\dag}_{-\mathbf{q}}+b_{\mathbf{q}}\right).
\label{Froehlich}
\end{eqnarray}
Here we use the Landau gauge for which the vector potential is given by $\mathbf{A}=(0, Hx, 0) $.
$a^{\dag}_\mathbf{k}$ and $a_\mathbf{k}$ are the creation and annihilation operators for electron states with
momentum $\mathbf{k}=(k_x,k_y)$ and energy $\varepsilon_0(\mathbf{k})$, while
$b^{\dag}_\mathbf{q}$, $b_\mathbf{q}$ are corresponding operators for phonon states with momentum
$\mathbf{q}=(q_x,q_y)$ and energy $\hbar\omega (\mathbf{q})$. The electron-phonon
coupling constant $g$ is assumed to be independent of $\mathbf{q}$. $A$ is the area of 2D crystal.

Let us now make the first approximation in the treatment of the Hamiltonian (\ref{Froehlich}). We replace
the phonon field by its mean-field value,
\begin{eqnarray}
g (b_\mathbf{q} + b_\mathbf{-q}^\dag) \rightarrow
g \left(\langle b_\mathbf{q} \rangle + \langle b_\mathbf{-q}^\dag
\rangle\right) = \delta_{\mathbf{q}, \mathbf{Q}} \sqrt{\textsl{A}} \Delta e^{i \Phi},
\label{Delta}
\end{eqnarray}
where the order parameter $\Delta e^{i \Phi}$ is the non-vanishing expectation value of
macroscopically occupied DW phonon mode.

The values of the order parameter amplitude $\Delta$ and the modulus of the DW momentum
$Q \equiv \mid\mathbf{Q}\mid$ will be determined later by the minimization of the total
energy of the system. Note that the direction of the wave vector is arbitrary. This
degeneracy follows from the assumed spatial isotropy of the band dispersion
 $\varepsilon_0(\mathbf{k})$, as well as of the phonon spectrum and of the
electron-phonon coupling. $\Phi$, the phase of the order parameter, is also arbitrary.
The phase degeneracy is present as far as one does not take into account any possible
pinning mechanisms, e. g. the commensurability of the DW and the crystal lattice, or the effects
of various irregularities (like impurities, crystal edges, etc).

After the mean-field step (\ref{Delta}) the Hamiltonian (\ref{Froehlich}) reduces to
\begin{eqnarray}
\mathcal{H}=\frac{1}{2 m} \sum_{\mathbf{k}}a^{\dag}_\mathbf{k}\left[ k_x^2+\left(k_y+i \frac{e
\hbar H}{c}\frac{\partial}{\partial k_x}\right)^2  \right]a_\mathbf{k} \nonumber \\
+\sum_{\mathbf{k}}\left[ \Delta e^{i \Phi} a^{\dag}_{\mathbf{k+Q}}a_\mathbf{k}+ \Delta e^{-i
\Phi} a^{\dag}_{\mathbf{k-Q}}a_\mathbf{k}\right] + \frac{\textsl{A} \hbar \omega_{Q}}{2 g^{2}}
\Delta^{2},
\label{meanfieldH}
\end{eqnarray}
where $\omega_{Q} \equiv \omega(\textbf{q} = \textbf{Q})$. The comprehensive analysis of this Hamiltonian, i. e. of the band reconstruction due to the DW
order, in the case of vanishing magnetic field ($H=0$) is given in the work \cite{KBR}. For finite
magnetic fields it is more convenient to diagonalize the Hamiltonian by using the coordinate
representation of electron field,
\begin{eqnarray}
\hat{\Psi}(\mathbf{r})=\sum_{\mathbf{k}}
a_\mathbf{k}\exp\{i \mathbf{kr}/\hbar\}.
\label{Psi}
\end{eqnarray}
The Hamiltonian (\ref{meanfieldH}) then reads
\begin{eqnarray}
\mathcal{H}&=&\frac{1}{2 m} \int \hat{\Psi}^\dag(\mathbf{r})\Big\{-\hbar^2
\frac{\partial^2}{\partial
x^2} + \Big(- i\hbar \frac{\partial}{\partial y}-\frac{eHx}{c}\Big)^2\nonumber \\
&+& 2 \Delta \cos\left(\mathbf{Q r}/\hbar+\Phi\right)\Big\}\hat{\Psi}(\mathbf{r}) +\frac{\textsl{A}
\hbar \omega_{Q}}{2 g^{2}} \Delta^{2}.
\label{coordinaterepH}
\end{eqnarray}
In the next Section we show how in the presence of both, finite DW modulation and finite magnetic
field, topological reconstruction of the Fermi surface takes place, changing electron dynamics
into the one of a peculiar magnetic breakdown type.

\section{Electron dynamics under magnetic breakdown conditions}

In this Section we assume strong magnetic fields, characterized by the regime
$\omega_0\ll \omega_H  \ll \varepsilon_F/\hbar$, where $\omega_0$ is the electron
relaxation frequency, and $\omega_H=e H/mc$ is the cyclotron frequency.
As one sees from Eq.(\ref{coordinaterepH}), the
electron states are the solutions of the effective one-electron Hamiltonian
\begin{eqnarray}
\mathcal{H}_{b}=\frac{1}{2 m} \Big\{-\hbar^2 \frac{\partial^2}{\partial x^2} + \Big(- i\hbar
\frac{\partial}{\partial y}-\frac{eHx}{c}\Big)^2 \Big\} + V(x),
 \label{BandH}
\end{eqnarray}
where $V(x)$ is the potential associated to the uniaxial DW charge modulation in the
$x$-direction,
\begin{eqnarray}
V(x)= 2\Delta \cos(Q x/\hbar).
\label{potential}
\end{eqnarray}
Here the free phase of the DW order parameter is chosen to be $\Phi=0$.

In the momentum representation, far from small regions were the MB takes
place (below we define these regions more precisely), the dynamics of electrons is
semiclassical. For further considerations it is convenient to use, like in Ref.\cite{KBR},
the new origin in the momentum space coinciding with the touching point due to the finite DW,
with corresponding electron momentum components $p_{x} = k_{x} + Q/2, p_{y} = k_{y}$, and the first Brillouin zone defined by $-Q/2 \leq p_{x} \leq Q/2$ (see Fig.\ref{trajectories} in which these coordinates are already used).
Then the Hamiltonian (\ref{BandH}) can be replaced by the Lifshitz - Onsager
equation \cite{LifshitzKosevich,Onsager}
\begin{eqnarray}
\varepsilon \left(p_x \pm Q/2, P_{y 0}-i \sigma \frac{d}{d p_x }\right)G(p_x,P_{y 0}) =\varepsilon
G(p_x,P_{y 0}), \nonumber \\
\varepsilon\left(p_x \pm Q/2, p_y\right)=\varepsilon,\,\,\,\,   \label{LOHamiltonian}
\end{eqnarray}
for wave functions $G(p_x,p_y)$, where $\varepsilon \left(p_x\pm Q/2, p_y\right)$ is the band dispersion
law of the initial system
in the absence of magnetic field $(H=0)$, with its center $p_{x} = p_{y} = 0$
in the origin of the coordinate system. Here $P_{y 0}$ is the conserved y-component of the
generalized momentum, $\sigma =\hbar^2/l_c^2 = e \hbar H/c$ is the magnetic area in the momentum
space, and $l_c =\sqrt{\hbar c/e H}$ is the magnetic length. The second line in
Eq.(\ref{LOHamiltonian}) is the energy conservation law that defines the semiclassical
electron trajectories in the momentum space shown in Fig.\ref{trajectories}.

The solution of Eq.(\ref{LOHamiltonian}) must satisfy the periodic boundary condition
$G(p_x,P_{y 0}) =G(p_x+Q,P_{y 0}) $ where the DW wave number $Q$ here appears as the
period of the chain in the $p_x$ direction.

The semiclassical solution of Eq.(\ref{LOHamiltonian}) reads
\begin{eqnarray}
G^{(J)}_{l} = \frac{C^{(J)}_{l}}{\sqrt{|v_l|}}\exp \Big\{ -\frac{i}{\sigma} \int_0^{p_x}
\Big(p_y^{(l,J)}(p_x^{\prime}) -P_{y0}\Big)d p_x^{\prime} \Big\}  \label{SemiclassFunction}
\end{eqnarray}
where $v_l= \partial \varepsilon/\partial p_y $ at $p_y = p_y^{(l)}(p_x)$ with indices
$l=1,2$ denoting lower and upper trajectories. Indices $J=I,II$ denote semiclassical
region, inside which the integration over $p_x^{\prime}$ is taken, that is left $(I)$ and right $(II)$
side of the reciprocal space with respect to the MB region, i. e. to
the origin defined by $p_{x} = 0$, as shown in Fig.\ref{trajectories}(b). The dependence
$p_y^{(l)}(p_x)$ is determined by the equality
$\varepsilon(p_x,p_y) =\varepsilon$ in Eq. (\ref{LOHamiltonian}).

These semiclassical considerations fail in narrow regions of the $p$-space in which semiclassical
trajectories closely approach each other. Then one has to take into account the MB - quantum tunneling
between the neighboring trajectories \cite{Falicov, Kaganov}. Dynamics of electrons
in these MB regions of ${\bf p}$-space is governed by the set of two equations \cite{SK,Slutskin}
\begin{eqnarray}
\Big[\varepsilon \left(p_x+Q/2, P_{y 0}-i \sigma \frac{d}{d p_x }\right) -\varepsilon \Big]G_1 + \Delta G_2=0,
 \nonumber \\
\Big[\varepsilon \left(p_x-Q/2, P_{y 0}-i \sigma \frac{d}{d p_x }\right) -\varepsilon \Big]G_2 + \Delta G_1=0,
\label{MBset}
\end{eqnarray}
where $\Delta$ is the amplitude of the DW order parameter introduced in Eq. (\ref{meanfieldH}).
Note that at $H=0$ the zeros of the determinant of this system of equations
give the dispersion law of the reconstructed bands
\begin{eqnarray}
\varepsilon_{\pm}(p_x,p_y) = \frac{1}{2}
\Big [\varepsilon(p_x+\frac{Q}{2},p_y) +\varepsilon(p_x-\frac{Q}{2},p_y) \nonumber \\
\pm \sqrt{\left( \varepsilon(p_x+\frac{Q}{2},p_y) -\varepsilon(p_x-\frac{Q}{2},p_y) \right)^2 +
4 \Delta^2}\Big].
\label{ReconstrH0}
\end{eqnarray}

In order to find the solution for finite magnetic fields, we note that
$p_x, p_y \ll p_F$ ($p_F$ is the Fermi momentum)
in the region of the MB, and hence the dispersion functions
$\varepsilon \left(p_x \pm Q/2, P_{y 0}-i \sigma \frac{d}{d p_x }\right)$ in the above set
of equations may be expanded in both their arguments.
Eqs. (\ref{MBset}) can be then solved
inside the MB region without using the semiclassical approximation. For magnetic fields
$\hbar \omega_H \ll \varepsilon_F $, the region in which the solution of this expanded set
is valid overlaps with the regions in which the semiclassical solutions
Eq.(\ref{SemiclassFunction}) are valid. Matching two sets of solutions, and taking into
account the above-mentioned periodic boundary condition, one finally gets the wave functions
and the dispersion equation for electrons under magnetic breakdown conditions.

As it was shown in our previous papers, in both cases of the initially open \cite{EPJB} and
closed \cite{PhysicaB} Fermi surfaces the dependence of the tunneling probability $t^2$
for MB between neighboring semiclassical trajectories on the DW order parameter $\Delta$,
the magnetic field $H$ and the Fermi energy $\varepsilon_F$ is qualitatively different for
two ranges of values of the DW momentum $Q$.

If $Q$ is small, so that the overlapping of the initial closed orbits is large, both velocities
in the above-mentioned expansion are large ($|v_x(0,0)|, |v_y(0,0)| \sim v_F$, where $v_F = p_{F}/m$ is the
Fermi velocity) and the tunneling
probability is given by the conventional Blount formula \cite{Blount,SK,Slutskin}
\begin{eqnarray}
t_{B}^2=1-\exp\{-\frac{\Delta^2}{\hbar \omega_H m |v_x^{(0)} v_y^{(0)}|}\}
\label{BlountEquation}
\end{eqnarray}
where $v_x^{(0)}$ and $v_y^{(0)}$ are the velocities at the crossing points of the initial
Fermi surfaces.

If $Q$ is close to $2p_F$, the initial Fermi surfaces nearly touch each other, one of the above-mentioned
velocities is close to zero (that is the electron system is close to the Lifshitz
$2\frac{1}{2}$ transition\cite{Lifshitz}), and  the tunneling probability is given
by\cite{Vorontsov,EPJB,PhysicaB,Fortin}
\begin{eqnarray}
t^2 =1-\exp\{-\frac{|A|^2 \Delta^2}{(\hbar \omega_H)^{4/3}
(\varepsilon_F)^{2/3}} \}, \nonumber \\
A (\varepsilon ; Q)=
2^{2/3}\pi \textrm{Ai}\left[2^{2/3}\frac{\varepsilon_Q-\varepsilon }{(\hbar
\omega_H)^{2/3}(\varepsilon_F)^{1/3}} \right],
\label{t}
\end{eqnarray}
where $\textrm{Ai}(x)$ is the Airy function and $\varepsilon_Q=(Q/2)^2/2m$.

As was shown in Refs.\cite{Vorontsov,EPJB,PhysicaB,Fortin}, solution of Eq.(\ref{MBset}) for
the case of our interest $Q \approx 2p_F$ can be presented by the integral
\begin{eqnarray}
G_{1,2}(p_x) \propto \int_{-\infty}^{\infty}g_{1,2}(\xi)\exp\Big\{p_x
\frac{(2 m v_x^{(0)})^{1/3}}{\sigma^{2/3}} \xi\Big\}d\xi,
\label{G}
\end{eqnarray}
in which $g_{1,2}(\xi)$ is a smooth function with the characteristic interval of variation
$\delta \xi \sim 1$. The wave functions $G_{1,2}(p_x)$ are of the semiclassical
character at $p_F \gg |p_x | \gg \sigma^{2/3}/(2 m v_x^{(0)} )^{1/3}$. Therefore, the regions in
which the matching can be done is near
$|p_x^{(match)} | = \sigma^{2/3}/(2 m v_x^{(0)})^{1/3}$. If $|p_x^{(match)} | \ll \Delta /v_F$
(see Eq.(\ref{ReconstrH0})), one may neglect the size of the MB region in comparison
to all characteristic parameters of the reconstructed spectrum. Accepting this inequality
and matching the wave functions with the usage of Eq.(\ref{t}) and the periodic boundary
condition, one finds the dispersion equation of electrons. Using this dispersion equation
one finally finds the density of states
(DOS) \cite{PhysicaB}
\begin{eqnarray}
\nu(\varepsilon) =\frac{2 S^{\prime}}{(2 \pi \hbar)^2} \frac{|\sin(S/2\sigma)| \Theta\big[t^2-
\cos^2(S/2\sigma)\big]}{\sqrt{t^2- \cos^2(S/2\sigma)}}.
\label{DOSH}
\end{eqnarray}
Here $ S^{\prime} \equiv d S(\varepsilon) / d \varepsilon$ and $ S(\varepsilon)$ is the area of
the periodic chain inside one of its periods (see Fig.\ref{trajectories}(b)). As one can see
from Eq.(\ref{DOSH}), the spectrum consists of Landau bands, of the width
$\sim t^2\hbar \omega_H$, centered around the discrete Landau levels
$E_n =\hbar \omega_H (n +1/2)$, and gaps of the width $\sim (1-t^2)\hbar \omega_H$
between them. Note that from Eq.(\ref{t}) it follows that the characteristic scale for the
cyclotron frequency $\omega_H$ appears to be $(\Delta/\varepsilon_{F})^{3/2}\varepsilon_{F}$, so that
the whole range of the strengths of magnetic field, including the asymptotic regimes of weak and strong
magnetic fields, is physically relevant and attainable. In particular, in the limit
$H \rightarrow \infty$ the MB probability $t^{2} \rightarrow 0$, electrons move along closed orbits and,
according to Eq.(\ref{DOSH}), $\nu(\varepsilon)$ goes to the conventional DOS of the electrons on closed
orbits under quantizing magnetic field. In the opposite limit of extremely weak magnetic fields the MB
probability $t^{2} \rightarrow 1$, electrons under the magnetic field move along open trajectories
(see Fig.1(b)) and $\nu(\varepsilon)$ goes to DOS of electrons in the absence of magnetic field.

The result (\ref{DOSH}) for $\nu(\varepsilon)$ is of central importance for the considerations
that follow. In particular, in the next section we calculate the electron density
\begin{eqnarray}
n(\varepsilon_F,\Delta,
Q)=\int_0^{\varepsilon_F} \nu(\varepsilon)d\varepsilon,
\label{elnumber}
\end{eqnarray}
and the electron band energy%
\begin{eqnarray}
E_b(\varepsilon_F,\Delta, Q) =
\int_0^{\varepsilon_F} \varepsilon \nu(\varepsilon)d\varepsilon
\label{bandenergy}
\end{eqnarray}
at temperature $T=0$.

\section{Band energy}

Using Eq.(\ref{DOSH}), along with the change  of integration variable
in Eqs.(\ref{elnumber}, \ref{bandenergy}) from $\varepsilon$ to $\varphi$,
\begin{eqnarray}
\frac{S(\varepsilon)}{2 \sigma} =\varphi,
\label{vchange}
\end{eqnarray}
one finds the electron density and the band energy as follows:
\begin{eqnarray}
n(\varepsilon_F)=\frac{4  \sigma}{(2 \pi \hbar)^2}\int_0^{\frac{S(\varepsilon_F)}{2 \sigma}}
\frac{|\sin\varphi|\Theta(t^2(\varphi)-\cos^2\varphi)}{\sqrt{t^2(\varphi)-\cos^2\varphi}}
d\varphi,\hspace{0.2cm}
\label{electrnumber2}
\end{eqnarray}
\begin{eqnarray}
E_b=\frac{4  \sigma}{(2 \pi \hbar)^2}  \hspace{6.2cm}  \nonumber \\
\times  \int_0^{\frac{S(\varepsilon_F)}{2 \sigma}}
\frac{\varepsilon(\varphi) |\sin\varphi|\Theta(t^2(\varphi)-\cos^2\varphi)}
{\sqrt{t^2(\varphi)-\cos^2\varphi}}d\varphi, \hspace{0.2cm}
\label{bandenergy2}
\end{eqnarray}
with $\varepsilon(\varphi) $ being the solution of Eq.(\ref{vchange}).

In Eqs.(\ref{electrnumber2},\ref{bandenergy2}), $\varepsilon_F$ is the Fermi energy of the
reconstructed system under magnetic field $H$, which is linked to the Fermi energy
$\bar{\varepsilon}_F$ of the same system at $H=0$ by the condition of conservation of the
electron number
\begin{eqnarray}
n(\varepsilon_F)=\frac{S(\bar{\varepsilon}_F)}{(2\pi \hbar)^2}.
\label{conservlaw}
\end{eqnarray}
On the other hand, $\bar{\varepsilon}_F$ is determined by the Fermi energy $\varepsilon_{F0}$
of the initial, unreconstructed system at $H=0$ through the condition of conservation of the number of
electrons at $H=0$,
\begin{eqnarray}
S(\bar{\varepsilon}_F, Q)= S_0(\varepsilon_{F0}),
\label{conservlaw2}
\end{eqnarray}
where $S_0(\varepsilon_{F0}) =\pi p_{F0}^2$ is the area of the initially closed Fermi surface
(see Fig.\ref{trajectories}(a)).

Carrying out the integration in Eq.(\ref{electrnumber2}) and taking into account the condition
(\ref{conservlaw2}), one arrives at the equation that defines the new Fermi energy of reconstructed
system in magnetic field (for details see Appendix \ref{AppendixElectronNumber}),
\begin{eqnarray}
\cos\left(\pi \delta n_F\right)=t\cos\left(\pi \delta n_{F0}\right).
\label{FSequation}
\end{eqnarray}
Here $\delta n_F$ and $\delta n_{F0}$ are the fractional parts of the ratios
\begin{eqnarray}
\frac{S(\varepsilon_F)}{2\pi \sigma} =N_F + \delta n_F, \nonumber \\
\frac{S_0(\varepsilon_{F0})}{2\pi  \sigma} =N_F
+ \delta n_{F0}
\label{FEcoupling}
\end{eqnarray}
respectively, while $N_F$ is integer.
Here and below we define
$0 \leq \delta n_F, \delta n_{F0} < 1$, and hence
$\varepsilon_{F0}$ coincides with a discrete Landau level if $\delta n_{F0}=1/2$ and it is the
middle between them if $\delta n_{F0} =0$ (the same is valid for $\varepsilon_F$).

Note that at extremely large magnetic fields one has $t \rightarrow 0$, with electrons
moving along closed orbits. In this limit $\delta n_F =1/2 $ at any value of $\delta n_{F0}$,
that is the Fermi energy of free electrons under magnetic field always coincides with one of the
Landau levels at any value of the filling factor.

With the Eq.(\ref{electrnumber2}) and the condition (\ref{FSequation}) taken into account,
the expression (\ref{bandenergy2}) for the band energy per unit area reduces
(details of the calculation are given in the Appendix \ref{AppendixBandEnergy}) to the following
convenient form
\begin{eqnarray}
E_b= E_b^{(H=0)} + E_b^{(H)}.
\label{Eband}
\end{eqnarray}

The first term on the right-hand side of Eq.(\ref{Eband}) is the band energy of the reconstructed
system in the absence of magnetic field \cite{KBR}
\begin{eqnarray}
E_b^{(H=0)}(Q, \Delta)=  \frac{4}{(2 \pi \hbar)^2m}\int_0^{\frac{Q}{2}} d p_x\int_0^{p_y^{(F)}}dp_y
\nonumber \\
\times \Big[\frac{Q}{2}^2 +p_y^2 +p_x^2 - \sqrt{(Q p_x)^2+(2m \Delta)^2}\Big],
\label{Eband0}
\end{eqnarray}
where
\begin{eqnarray}
p_y^{(F)}   =\Big\{  2m \varepsilon_F   -\frac{Q}{2}^2 - p_x^2+  \sqrt{(Q p_x)^2+(2m \Delta)^2}
\Big\}^{\frac{1}{2}}.\hspace{0.2cm}
\label{py}
\end{eqnarray}
Detailed analysis of this energy and conditions of the stabilization of the DW at $H=0$ is
presented in Ref.\cite{KBR}.

The second term in Eq.(\ref{Eband}) is the electron "magnetic"
energy, including the contribution of MB that takes place in small regions in the vicinity of
points of the closest approach of the two open trajectories,
\begin{flalign}
E_b^{(H)} = \nu_0 \frac{(\hbar \omega_H)^2}{2} \Big\{ - \delta n_{F0}^2 \hspace{3.0cm} \nonumber \\
+\frac{2}{\pi^2} \int_{\cos(\pi \delta n_{F0})}^1
\frac{\arccos\left(t\zeta\right)}{\sqrt{1-\zeta^2}} d\zeta \Big\}.
\label{EbandH}
\end{flalign}
Here $\nu_0 =4 \pi m /(2 \pi \hbar)^2$
is the band density of states of the unreconstructed system at $H=0$.

When magnetic field is rather small one has $t \approx 1$ (see Eq.(\ref{t})), i. e.
the MB is absent, and electrons move along open trajectories. As it follows from
Eqs.(\ref{FSequation},\ref{Eband}) the electron "magnetic" energy then tends to zero.
Therefore, one has
\begin{eqnarray}
E_b=E_b^{(H=0)}(Q, \Delta),
\label{Ebandt1}
\end{eqnarray}
that is the band energy and the Fermi energy (see Eq.(\ref{conservlaw2})) are the same as in
the absence of magnetic field.

In the other limiting case of a strong magnetic field, MB is strong ($t \rightarrow 0$),
so that the electrons move along closed orbits and, according to
Eqs.(\ref{FSequation},\ref{Eband}), one has $\delta n_F = \frac{1}{2}$ and
\begin{eqnarray}
E_b=E_b^{(H=0)}(Q, \Delta)
+  \nu_0 \frac{(\hbar \omega_H)^2}{2} \delta n_{F0} \left(1-\delta n_{F0}\right).
\label{Eband2}
\end{eqnarray}

On the other hand, taking $\Delta =0$ in Eq.(\ref{Eband}) one finds the band
energy of the initial, unreconstructed system in the absence of the DW as follows:
$\delta n_F= \frac{1}{2}$ and
\begin{eqnarray}
E_b^{(\Delta=0)}= \frac{1}{2}\nu_0\varepsilon_{F0}^2 \Big\{ 1+ \left(\frac{\hbar \omega_H}{\varepsilon_{F0} }\right)^2
\delta n_{F0} \left(1-\delta n_{F0}\right) \Big\}. \hspace{0.2cm}
\label{Eband3}
\end{eqnarray}

Here $\frac{1}{2}\nu_0\varepsilon_{F0}^2$ is the initial band energy in
the absence of the DW and of magnetic field. Therefore, in this limit, in both cases of reconstructed
and unreconstructed systems (Eqs.(\ref{Eband2}, \ref{Eband3})) the Fermi energy, $\varepsilon_{F}$,
at $ H\neq 0$ coincides with one of the discrete Landau levels independently of the value of the
Fermi energy $\varepsilon_{F0}$ at $H=0$, while the band energy of electrons localized by the
magnetic field (Landau electrons) is always greater than, or equal to the band energy at
$H=0$.

Subtracting the initial band energy Eq.(\ref{Eband3}) from Eq.(\ref{Eband}), one finds the
decrease of the total band energy per unit area as follows:
\begin{eqnarray}
\Delta E_b=\Delta E_b^{(H=0)}(Q, \Delta) +  \hspace{4.5cm}\nonumber \\
\nu_0 \frac{(\hbar \omega_H)^2}{2}
\Big\{-\delta n_{F0}+ \frac{2}{\pi^2} \int_{\cos(\pi \delta n_{F0})}^1
\frac{\arccos\left(t\zeta\right)}{\sqrt{1-\zeta^2}} d\zeta  \Big\}. \hspace{0.2cm}
\label{Ebandt2}
\end{eqnarray}

\begin{figure}
\centerline{\includegraphics[width=\columnwidth]{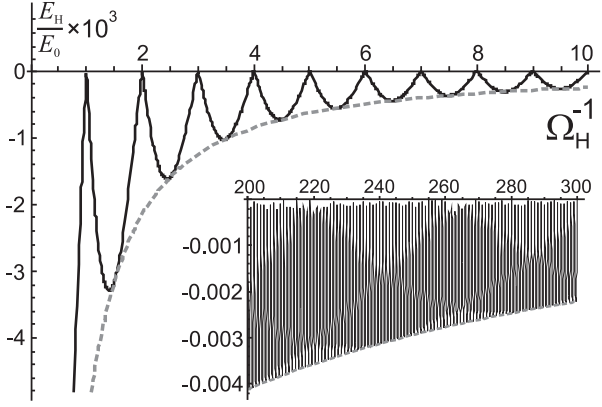}}
\caption{
De Haas - van Alphen type of oscillations of the magnetic energy gain (given by the
second term in Eq. (\ref{Ebandt2})), with respect to the inverse magnetic field
$\Omega_H^{-1} \equiv \varepsilon_{F0}/\hbar \omega_H $. The dotted curve is the
envelope showing the maximal magnetic energy gain obtained for $\delta n_{F0}=1/2$.
The inset shows the experimentally relevant interval $30T \leq H \leq 50T$, for the
choice of parameters $\Delta/\varepsilon_{F0}=0.01$, and such value of $Q$ that maximizes
$t^2$ in Eq.(\ref{t}).}
\label{oscillations}
\end{figure}

The first term on the right-hand side of Eq.(\ref{Ebandt2}) is the energy gain of the
reconstructed system with the DW in the absence of magnetic field, $H=0$. Its Fermi energy
$\bar{\varepsilon}_F$ is determined by Eq.(\ref{conservlaw2}). This energy was derived
and in details analyzed in our previous paper \cite{KBR}, in which it was proved that the DW energy is optimal if
$\bar{\varepsilon}_F = \varepsilon_{F0}$. The second term is the electron
magnetic energy gain which is responsible for the de Haas-van Alphen oscillations modified
by magnetic breakdown \cite{coment}. This magnetic energy as a function of the magnetic field is shown
in Fig.\ref{oscillations}.

%%%%%%%%%%%%%%%%%%%%%%%%%%
%%%%%%%%%%%%%%%%%%%%%%%%%%%%

\section{Stability of the DW order}

Besides the band energy (\ref{Ebandt2}), the complete condensation energy $E_{DW}$ includes also
the contribution from the periodic lattice deformation, given by the corresponding mean-field term
in the Hamiltonian (\ref{meanfieldH}). For the further analysis, more precisely for the minimization
of $E_{DW}$ with respect to the DW momentum $Q$ and the order parameter $\Delta$, it is convenient
to introduce dimensionless quantities
\begin{eqnarray}
q \equiv \frac{Q}{2p_{F0}}, \hspace{1.0cm}
\delta \equiv \frac{\Delta}{\varepsilon_{F0}}, \nonumber \\
\Omega_H \equiv \frac{\hbar \omega_H}{\varepsilon_{F0}}, \hspace{1.0cm}
\lambda \equiv \frac{m}{\pi \hbar^2}\frac{g^2}{2 \hbar \omega_Q},
\label{scaling}
\end{eqnarray}
for the DW wave vector, DW order parameter, magnetic field, and coupling constant respectively.
$\lambda$ is, like $g$, assumed to be $\textbf{Q}$-independent, i. e. we do not take into
account the presumably smooth dependence of $\omega_{Q}$ on $\textbf{Q}$ in the narrow range
of values of $\textbf{Q}$ related to the touching instability from Fig.(\ref{trajectories}).
Written in terms of quantities (\ref{scaling}), the total condensation energy $E_{DW}$, scaled by
the bare band energy $E_0=\frac{1}{2}\nu_0\varepsilon_{F0}^2$ introduced
after Eq.(\ref{Eband3}), reads
\begin{eqnarray}
\frac{E_{DW}(\Omega_H,q,\delta)}{E_{0}} = \widetilde{E}_{DW}^0 + \widetilde{E}_{DW}^H,
\label{EDWscaled}
\end{eqnarray}
were the dimensionless (scaled) partial condensation energy terms in expression (\ref{EDWscaled}) are
\begin{eqnarray}
\widetilde{E}_{DW}^0 (q,\delta) = 1- \frac{16}{3\pi} \int_{0}^{q} \Big( 1-q^2-x^2 +  \hspace{2.2cm} \nonumber \\
 \sqrt{(2qx)^2 + \delta^2} \Big)^{\frac{3}{2}} dx +\frac{\delta^2}{\lambda} \nonumber \\
\widetilde{E}_{DW}^H (\Omega_H,q,\delta) = \Omega_H^2 \Big\{-\delta n_{F0} \hspace{3.5cm} \nonumber \\
+ \frac{2}{\pi^2} \int_{\cos(\pi \delta n_{F0})}^1 \frac{\arccos\left(|t|\zeta\right)}{\sqrt{1-\zeta^2}} d\zeta  \Big\},\hspace{0.2cm}
\label{Epartscaled}
\end{eqnarray}
where the last term in the first expression is due to the lattice deformation.

Let us by $q_{m}$ denote the optimal value of the DW wave number for which the condensation
energy (\ref{EDWscaled}) has the minimum. To simplify its determination we note at first that
both $q$-dependent condensation energy contributions in (\ref{Epartscaled}) have their minima
at the wave numbers in the range $Q \sim 2p_{F}$. Let us denote these minima by $q_{m}^{0}$ and
$q_{m}^{H}$ respectively. The former is explicitly given by\cite{KBR}
\begin{eqnarray}
q_m^{0}=
1-\frac{\delta}{2}+\frac{\delta^{3/2}}{2\sqrt{2}\pi}.
\label{qmH0}
\end{eqnarray}
As for the latter, the minimization of $\widetilde{E}_{DW}^H$ gives
\begin{eqnarray}
q_m^{H}=\sqrt{1+2^{-2/3}x_m \Omega_H^{2/3}},
\label{qmH}
\end{eqnarray}
where $x_m=-1.01879$ is the value that maximizes the Airy function in Eq.(\ref{t}), thus
providing the equilibrium value of the tunneling probability
\begin{eqnarray}
t_m^2 =1-\exp\{-2^{4/3}\pi^2 \textrm{Ai}^2(x_m)  \frac{\delta^2}{\Omega_H^{4/3}} \} \nonumber \\
=1-\exp\{-7.14  \frac{\delta^2}{\Omega_H^{4/3}} \}.
\label{tm}
\end{eqnarray}

Let us now approximate the expressions for $\widetilde{E}_{DW}^0$ and $\widetilde{E}_{DW}^H$ by the
respective quadratic expansions around $q_{m}^{0}$ and $q_{m}^{H}$:
\begin{eqnarray}
\widetilde{E}_{DW}^{0,H}(q) \approx \widetilde{E}_{DW}^{0,H}(q=q_{m}^{0,H}) + \frac{1}{2} \alpha_{0,H} (q - q_{m}^{0,H})^{2},
\label{EDWexp}
\end{eqnarray}
with corresponding minimal energies
\begin{eqnarray}
\widetilde{E}_{DW}^0 (q_m^{0})=\left( \frac{1}{\lambda}-\frac{1}{\lambda_c^{(H=0)}} \right) \delta^2+
\frac{1}{\pi}\delta^3,
\label{EDW0-0}
\end{eqnarray}
\begin{eqnarray}
\widetilde{E}_{DW}^H (q_m^{H})=\Omega_H^2 \Big\{-\delta n_{F0} \hspace{4.0cm} \nonumber \\
+ \frac{2}{\pi^2} \int_{\cos(\pi \delta n_{F0})}^1
\frac{\arccos\left(t_m\zeta\right)}{\sqrt{1-\zeta^2}} d\zeta  \Big\}, \hspace{0.2cm}
\label{EDWH-0}
\end{eqnarray}
and expansion coefficients
\begin{eqnarray}
\alpha_{0}=\frac{16}{\pi} \left( 1-\sqrt{2}+\frac{\pi}{\sqrt{2}} \right) \delta^{\frac{1}{2}},
\label{alpha0}
\end{eqnarray}
\begin{eqnarray}
\alpha_{H}=2^{\frac{2}{3}}32 x_m \textrm{Ai}^2(x_m) \frac{\delta^2}{\Omega_H^{2/3}}
\frac{1-t_m^2}{t_m^2} \ln{\sqrt{\frac{1-t_m}{1+t_m}}}.
\label{alphaH}
\end{eqnarray}
The DW condensation energy in the absence of magnetic field, taken at its optimal wave
vector, is given by Eq. (\ref{EDW0-0}) and has been elaborated in detail in the work\cite{KBR}.
There $\lambda_c^{(H=0)}=(1+2/\pi)^{-1} \approx 0.61$ is the critical value of the coupling
constant for the DW stabilization at $H=0$. Analogously, $\widetilde{E}_{DW}^H (q_m^{H})$
in Eq.(\ref{EDWH-0}) is the magnetic energy taken at its own optimal wave vector $q_m^{H}$.
The coefficient $\alpha_{H}$ in Eq.(\ref{alphaH}) is provided taking only the envelope of
magnetic energy oscillations into account, i.e. taking $\delta n_{F0}=1/2$ (see
Fig.\ref{oscillations}). The same manner will be used in the presentation of further results.

After inserting the expansions (\ref{EDWexp}) into the expression (\ref{EDWscaled}), and the
minimization of the latter with respect to $q$, one finally gets the optimal condensation energy
\begin{eqnarray}
\frac{E_{DW}}{E_0}=\widetilde{E}_{DW}^0(q_m^{0})+\widetilde{E}_{DW}^H(q_m^{H}) \hspace{3.0cm} \nonumber \\
+\frac{1}{2}\frac{\alpha_{0}\alpha_{H}}{\alpha_{0}+\alpha_{H}} (q_m^{0}-q_m^{H})^2,\hspace{1.2cm}
\label{EDWmin}
\end{eqnarray}
and the optimal DW wave vector
\begin{eqnarray}
q_{m} = \frac{\alpha_{0}q_{m}^{0}+\alpha_{H}q_{m}^{H}}{\alpha_{0}+\alpha_{H}}.
\label{qm}
\end{eqnarray}

\begin{figure}
\centerline{\includegraphics[width=\columnwidth]{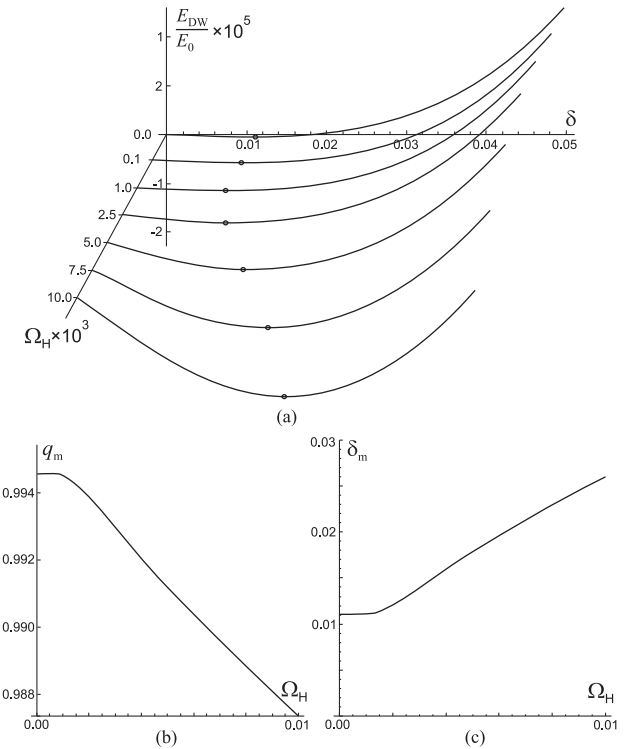}}
\caption{(a) The dependance of the DW condensation energy (\ref{EDWscaled}) on
the order parameter $\delta$ and the magnetic field $\Omega_H$, taken at the
optimal value of wave vector $q_m$ and for $\lambda=0.613$. The dependence of (b) optimal DW wave vector
$q_m$, and of (c) optimal DW order parameter $\delta_m$, on $\Omega_H$, for the same value of
$\lambda$.}
\label{energy}
\end{figure}

The dependance of the DW condensation energy (\ref{EDWmin}) on the order parameter and
magnetic field for $q = q_m$ is shown in Fig.\ref{energy}(a).
It is evident that by increasing magnetic field the minimum of $E_{DW}$
is lowering. This shows that magnetic field additionally strengthens stabilization of the DW.

The results of the numerical calculation of the dependence of the DW wave number $q_m$ and
of the DW amplitude $\delta$ on magnetic field for a given value of the coupling constant
$\lambda$ are shown in Figs.\ref{energy}(b, c). $q_m$ shows a very weak dependence on
magnetic field of the order of only few percent, while the order parameter increases
approximately three times within the same range of variation of magnetic field, corresponding to
the span of hundred Tesla.

%%%%%%%%%%%%%%%%%%%%%%%%%%
%%%%%%%%%%%%%%%%%%%%%%%%%%%%

\begin{figure}
\centerline{\includegraphics[width=\columnwidth]{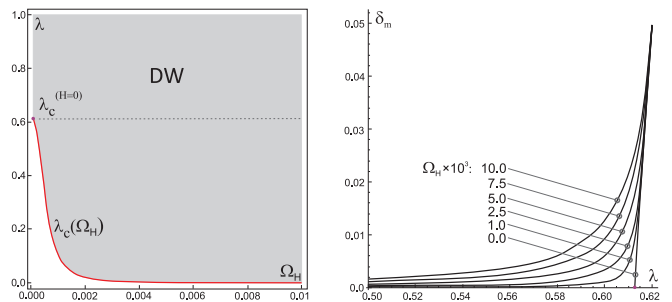}}
\caption{Phase diagram:
(a) The DW is present in the shaded domain of the
$(\Omega_H,\lambda)$ plane, with the curve $\lambda_c (\Omega_H)$ representing
the critical value of the coupling constant below which the DW order
parameter vanishes. The dashed line at $\lambda = \lambda_c^{H=0}$ is critical line
below which the DW order is possible only in finite magnetic field.
(b) DW order parameter $\delta_m$ \emph{vs} $\lambda$ for a series of values
of $\Omega_H$.}
\label{phasediagram}
\end{figure}

The above results lead us to the phase diagram shown in Fig.\ref{phasediagram}(a).
As already pointed out in the previous work \cite{KBR}, in the absence of magnetic field
the phase transition resulting in the DW and reconstructed electron band has the
characteristics of the quantum phase transition, which takes place only providing
$\lambda > \lambda_c^{(H=0)}$.

The present analysis indicates that regarding this transition a finite external magnetic field
strengthens the DW ordering. Firstly, as shown in Fig.\ref{energy}(a), it lowers the total energy of the DW state.
Secondly, Fig.\ref{energy}(c) shows that the magnitude of the order parameter $\delta$ increases as $H$ increases.
Furthermore, by switching the external magnetic field one introduces a qualitative change into the DW phase diagram.
Namely, as Fig.\ref{phasediagram}(a) shows, the domain of values of parameters $\Omega_H$ and $\lambda$ for which the DW is stable
extends towards lower and lower values of the coupling constant $\lambda$ as $\Omega_H$ increases.
Only below the critical curve $\lambda_c(\Omega_H)$ the ordering is suppressed, i. e. the order parameter $\delta$ vanishes
within the range of numerical imprecision. We see that such effective critical coupling $\lambda_c(\Omega_H)$
saturates to zero at a finite value of $H$, roughly estimated to be of the order of
few dozens Tesla. In other words, there is a range of values of the coupling constant in which
the DW order and the band reconstruction, although not possible at $H=0$, can be induced by
applying the external magnetic field. In Fig.\ref{phasediagram}(b) we show the equilibrium value of the DW order parameter
$\delta_m$ \emph{vs} $\lambda$ for a series of different values of magnetic field $\Omega_H$.

%%%%%%%%%%%%%%%%%%%%%%%%%%
%%%%%%%%%%%%%%%%%%%%%%%%%%%%

\section{Conclusion}

The results presented above show that a strong enough external magnetic field strengthens
the tendency of isotropic 2D electron band towards the band reconstruction, associated with
the formation of uniaxial DW that breaks the translational symmetry of the system.
The band reconstruction under magnetic field is the result of two opposing tendencies. From
one side, in the magnetic field open 1D trajectories are additionally energetically favored
with respect to the closed Landau quantized orbits in the initial non-reconstructed 2D band.
From the other side, the closeness of oppositely directed open trajectories from
Fig.\ref{trajectories} inevitably provokes a MB between them. Finite tunneling probability
between these trajectories causes a partial re-localization of band states, and thus acts
against the DW stability.

In the above analysis we use the well-established semiclassical approach of electron dynamics
in the magnetic field, together with the full quantum mechanical treatment of the MB. The obtained
band spectrum consists of an alternating sequence of narrow energy sub-bands and energy
sub-gaps \cite{Kaganov}, each sub-band being located around a Landau discrete level.

The knowledge of the associated density of states (\ref{DOSH}) opened the way towards the
calculation of the total condensation energy at zero temperature, given by Eq.(\ref{EDWscaled}).
The important outcome of this calculation is the additivity of the band energy gain. It has two
contributions, one due to the bare $H=0$ band reconstruction, and another due to the above-mentioned effects
of finite magnetic field. The former comes from redistribution of the density of states towards
lower energies in the reconstructed band \cite{KBR}. The latter comes, as already stated above, from the
delocalization of initially Landau-localized electrons, modified by the magnetic breakdown. It
is accompanied with quantum oscillations periodic in $1/H$, similar to the standard de
Haas - van Alphen effect, due to Landau level filling (see Fig.\ref{oscillations}).

The next important element which facilitates the further minimization of the condensation energy
(\ref{EDWscaled}) is the closeness of values of optimal wave numbers which minimize these
two contributions. Both are close to $2p_{F}/\hbar$ (see Fig.\ref{trajectories}), differing only
by few percents. The final optimal DW wave number, minimizing the total condensation energy, is
an adequate mean of these two, given by Eq.(\ref{qm}) and shown in Fig. \ref{energy}.

The central result of the whole analysis is the phase diagram shown in Fig. \ref{phasediagram},
from which one concludes that the external magnetic field strengthens the DW quantum phase
transition with respect to that stabilized at $H=0$. One measure of this strengthening is the
increase of the order parameter by approximately three times along the span of magnetic field
up to hundred Tesla. Simultaneously the optimal wave number decreases only by few percent in
the same domain, the reasons for this weak variation being already pointed out above. Finally,
the critical value of the coupling constant $\lambda$, beyond which quantum phase transition
takes place, decreases by increasing $H$, and quickly saturates towards zero at the field domain
of few dozens Tesla. The external magnetic field thus opens possibility for DW and band
reconstruction to appear in materials in which they are not present in zero field.

\textbf{\emph{Acknowledgement}}. This work was supported by the Croatian Science Foundation,
project IP-2016-06-2289, and by the QuantiXLie Centre of Excellence, a project cofinanced by
the Croatian Government and European Union through the European Regional Development Fund -
the Competitiveness and Cohesion Operational Programme (Grant KK.01.1.1.01.0004).

\bigskip

\begin{appendix}
\section{Calculations of the electron number
\label{AppendixElectronNumber}}

As $S(\varepsilon_F)/2\sigma =   \varepsilon_F/ \hbar \omega_H \gg 1$,  it is convenient
to re-write the upper integration limit of the first integral in Eq.(\ref{electrnumber2})
in the form $\pi (N_F + \delta n_F)$ (where $N_F$ and $0 \leq \delta n_F <1$ are the integer
and fractional parts of $S(\varepsilon_F)/2\sigma$ respectively, and present the integral as
a sum of integrals over two intervals, $(l \pi, (l+1)\pi )$ and $(0 \leq l \leq N_F-1)$.
After the change of variables $\varphi -l \pi\rightarrow \varphi$ one finds
\begin{eqnarray}
n= \frac{4 \sigma}{( 2\pi \hbar)^2}\sum_{l=0}^{N_F-1} \int_0^{\pi}
\frac{|\sin\varphi|\Theta(t_l^2-\cos^2\varphi)}{\sqrt{t_l^2-\cos^2\varphi}}
d\varphi \nonumber \\
+\int_0^{\pi \delta n_F}
\frac{|\sin\varphi|\Theta(t^2-\cos^2\varphi)}{\sqrt{t^2-\cos^2\varphi}}
d\varphi
\label{ElectrNumberAppendix2}
\end{eqnarray}
where $t_l=t(\varphi +l \pi)$ and $t=t(\varphi+ (N_F-1) \pi) \approx t(\varepsilon_F)$.

As one sees from Eq.(\ref{BlountEquation},\ref{t})  the characteristic variation interval of
$t(\varphi)$ is $\delta \varphi \sim (\varepsilon_F/\hbar \omega_H)^{1/3}\pi \gg \pi$ and
hence $t_l$ and $t$  may be considered as  constants inside each interval of the integrations
with high accuracy. Under  the latter condition, the integrals under the sum in
Eq.(\ref{ElectrNumberAppendix2}) are table integrals equal to $\pi$, and hence the electron
density reads
\begin{eqnarray}
n=\left[\pi N_F
+\int^1_{\cos\frac{\pi \delta n_F}{t}}
\frac{\Theta(1-\xi^2)}{\sqrt{1-\xi^2}}
d\xi \right]
\label{ElectrNumberAppendix3}
\end{eqnarray}
While writing the above equation we changed variable in the last integral in
Eq.(\ref{ElectrNumberAppendix2}) $\cos\varphi=t \xi$.

Inserting this result into the electron number conservation law given by
Eqs.(\ref{conservlaw},\ref{conservlaw2}) one gets the equation that couples the
Fermi energy of the metal with DW and the initial Fermi energy $\varepsilon_{F0}$ at
$H=0$:
\begin{eqnarray}
\pi \delta n_{F_0}=\int^1_{\cos\frac{\pi \delta n_F}{t}}
\frac{\Theta(1-\xi^2)}{\sqrt{1-\xi^2}} d\xi.
\label{AppendixCoservLaw1}
\end{eqnarray}
Here $\delta n_{F0}$ is the fractional part of $S(\varepsilon_{F0})/2 \sigma$. As $\delta
n_F <1$ by definition, the only solution of this equation is Eq.(\ref{FSequation}) of the main
text.

\section{Calculations of the band energy
\label{AppendixBandEnergy}}

After presenting the integral in Eq.(\ref{bandenergy2}) as a sum of integrals over
intervals $(l \pi, (l+1)\pi )$ and $(0 \leq l \leq N_F-1)$, and changing the variables
$\varphi -l \pi\rightarrow \varphi$ in the same way as in Appendix
\ref{AppendixElectronNumber}, the band energy reads
\begin{eqnarray}
E_b^{(H)}=E_b^{(1)}+E_b^{(2)},
 \label{AppendixBandSum}
\end{eqnarray}
where
\begin{eqnarray}
E_b^{(1)}=\frac{4 \sigma}{( 2\pi \hbar)^2}
\int_0^{\pi \delta n_F} \varepsilon \Big(\pi N_F+\varphi \Big) B\Big[\varphi ; t
\Big(\varepsilon_F \Big)\Big] d\varphi \hspace{0.2cm}
\label{EnergyAppendixE1}
\end{eqnarray}
and
\begin{eqnarray}
E_b^{(2)}=\frac{4 \sigma}{( 2\pi \hbar)^2} \hspace{5.2cm}\nonumber \\
\times\int_0^{\pi}d\varphi
\sum_{l=0}^{N_F-1}\varepsilon \Big(\pi l+\varphi \Big)B\Big[\phi;
t\Big(\varepsilon(l \pi +\varphi)\Big)\Big].
\label{EnergyAppendixE2}
\end{eqnarray}
Here
\begin{eqnarray}
B\Big[\varphi; t\Big(\varepsilon(\varphi)\Big)\Big]=
\frac{|\sin\varphi| \Theta\Big[ t [\varepsilon(\varphi) ]^2-\cos^2\varphi
\Big]}{\sqrt{t [\varepsilon(\varphi) ]^2-\cos^2\varphi}}.
\label{B}
\end{eqnarray}

\bigskip

\textbf{BI. Calculations of the integral in Eq.(\ref{EnergyAppendixE1})}

\bigskip

The dependence of energy on $\varphi$ in Eq.(\ref{EnergyAppendixE1}) is
determined by the equation
\begin{eqnarray}
\frac{S(\varepsilon)}{2 \sigma }= \pi N_F +\varphi.
\label{SApp1}
\end{eqnarray}
As the integration here goes in the vicinity of the Fermi energy in the range
$\lesssim \hbar \omega_H$, it is convenient to expand the area here as
$S(\varepsilon) \approx S(\bar{\varepsilon}_F) +\delta \varepsilon S^{\prime}(\bar{\epsilon}_F)$,
where $\delta \varepsilon = \varepsilon -\bar{ \varepsilon}_F$.

According to Ref. \cite{KBR}
\begin{eqnarray}
S(\varepsilon)=  \hspace{6.2cm}         \nonumber \\
4 \int_0^{Q/2}\sqrt{2 m \varepsilon - \frac{Q}{2}^2-p_x^2+\sqrt{(Qp_x)^2+(2 m \Delta)^2}}.
\label{S1}
\end{eqnarray}
Taking derivative with respect to $\varepsilon$ and changing variables
$p_x =p_F \xi$ one finds
\begin{eqnarray}
S(\varepsilon)=   S(\varepsilon_F)    \hspace{5.2cm}     \nonumber \\
+4 m \delta \varepsilon\int_0^1 \frac{d \xi}
{\sqrt{\delta - q^2-\xi^2+\sqrt{4 \xi^2+ \delta^2}}},
\label{S1}
\end{eqnarray}
where $\delta = \Delta/  \varepsilon_F$ and  $q$  in the last term is taken equal
to $q = p_F (1-\delta/2)$ that presents the optimal DW vector in the lowest
approximation in $\delta , \;\; \hbar \omega_H/\varepsilon_F \ll 1$.

Changing variables $\zeta^2 = 4 \xi^2 + \delta^2$ and performing integration
one finds the needed expansion as follows:
\begin{eqnarray}
S(\varepsilon)=   S(\bar{\varepsilon}_F)
+2 \pi m( \varepsilon - \bar{\varepsilon}_F)
\label{S1}
\end{eqnarray}

Rewriting Eq.(\ref{SApp1}) as $S(( \varepsilon) =2 \sigma [\pi (N_F + \delta
\bar{n}_F) -\pi \delta \bar{n}_F + \varphi]=S( \bar{\varepsilon}_F)+2\sigma
(\varphi - \pi \delta \bar{n}_F) $ with the usage of Eq.(\ref{S1}) one finds
\begin{eqnarray}
\varepsilon(\varphi +  \pi  N_F)=\bar{ \varepsilon}_F + \frac{\sigma}{\pi m}
\left(\varphi - \pi \delta \bar{n}_F\right),
\label{energyexpansion1}
\end{eqnarray}
where $\delta \bar{n}_F$ is the fractional part of the ratio
\begin{eqnarray}
\frac{S(\bar{\varepsilon}_{F})}{2\pi \sigma}
= N_F + \delta \bar{ n}_{F},
\label{FEcouplingbis}
\end{eqnarray}
while $N_F$ is integer. Note that according to Eq.(\ref{conservlaw2}) one has
$\delta \bar{n}_F= \delta  n_{F0}$.

Inserting the above equation into Eq.(\ref{EnergyAppendixE1}), and performing
the integration with respect to $\varphi$, one finds
\begin{eqnarray}
E_b^{(1)}=\frac{ \sigma}{( 2\pi \hbar)^2}\Big\{
\pi \bar{ \varepsilon}_F \delta \bar{n}_F     \hspace{3.5cm} \nonumber \\
+\frac{\sigma}{\pi m}\Big[ \int_{\cos{(\pi \delta \bar{n}_F)}}^{1}
\frac{\arccos{(t\zeta)}}{\sqrt{1-\zeta^2}} d\zeta-(\pi \delta \bar{n}_F)^2\Big] \Big\}.
\label{EnergyAppendixE1final}
\end{eqnarray}

\bigskip

\textbf{B.II. Calculations of $E_b^{(2)}$}

\bigskip

The sum in Eq.(\ref{EnergyAppendixE2})
\begin{eqnarray}
A =\sum_{l=0}^{N_F-1}\varepsilon \Big(\pi l+\varphi \Big)B\Big[\varphi;
t\Big(\varepsilon(l \pi +\varphi)
\Big)\Big]
\label{B1}
\end{eqnarray}
may be presented as follows:
\begin{eqnarray}
A =\frac{1}{\pi}\sum_{k=-\infty}^{\infty}  \int_{-\pi/2}^{(N_F-1/2)\pi}d x
\exp\{i2k x -2|k|\eta\}\nonumber \\
\times \varepsilon \Big(x+\varphi \Big)B\Big[\varphi;
t\Big(\varepsilon(x+\varphi)
\Big)\Big].
\label{B2}
\end{eqnarray}
While writing this equation we used the equality
\begin{eqnarray}
\sum_{l=-\infty}^{\infty}  \delta [x-l \pi] =\frac{1}{\pi}
\sum_{k=-\infty}^{\infty} \exp\{i2k x -2|k|\eta\},
\label{deltafunction}
\end{eqnarray}
where $\eta \rightarrow 0$.

Changing the integration variables
\begin{eqnarray}
\frac{S(\varepsilon)}{2 \sigma} =x +\varphi
\label{variablechange}
\end{eqnarray}
one finds
\begin{eqnarray}
A=\frac{1}{2 \pi \sigma} \int_{\varepsilon_1}^{\varepsilon_2}
\varepsilon S^{\prime}(\varepsilon)
B\Big[\varphi; t(\varepsilon)\Big] d \varepsilon
+ \sum_{k=1}^{\infty} \exp\{ -2 k \eta\}\nonumber \\
\times \int_{\varepsilon_1}^{\varepsilon_2}
\varepsilon S^{\prime}(\varepsilon)
B\Big[\varphi; t(\varepsilon)\Big] \Big( e^{ i2k\left(\frac{S(\varepsilon)}
{2 \sigma}-\varphi\right)}   +c.c.      \Big), \hspace{0.2cm}
\label{A1}
\end{eqnarray}
where $\varepsilon_{1,2}$ are defined by the equations
\begin{eqnarray}
\frac{S(\varepsilon_1)}{2 \sigma} =-\frac{\pi}{2} +\varphi;
\hspace{0.3cm}\frac{S(\varepsilon_2)}{2 \sigma} =
(N_F-\frac{1}{2})\pi +\varphi.
\label{variablechange2}
\end{eqnarray}

The main contributions to the integrals under summation on the right-hand side
of the above equation come from the ends of integration interval $\varepsilon_1$
and $\varepsilon_2$ because $S/2 \sigma \gg 1$, and the exponents there are fast
oscillating functions, while $S^{\prime} \neq 0$ (hence there is no saddle point).
Expanding the functions in the integrand in the vicinity of the ending points
$\varepsilon_{1,2}$ and carrying out the integration, one finds that the integral
under the summation sign is equal to zero at any $k\neq 0$, and hence the first
term on the right-hand side of Eq.(\ref{A1}) only remains. Inserting it into
Eq.(\ref{EnergyAppendixE2}) one finds first term on the right-hand side of the band
energy as follows:
 \begin{eqnarray}
E_b^{(2)}=\frac{2}{( 2\pi \hbar)^2 \pi}
\int_0^{\pi}d\varphi \int_{\varepsilon_1}^{\varepsilon_2} d \varepsilon
\varepsilon S^{\prime}(\varepsilon)
B\Big[\varphi; t(\varepsilon)\Big].
\label{EnergyAppendixB2A}
\end{eqnarray}

Further on, changing variables in the integral with respect to the energy
\begin{eqnarray}
\frac{S(\varepsilon)}{2 \sigma} =\varphi_1+\varphi -\frac{\pi}{2}
\label{variablechange3}
\end{eqnarray}
one finds
\begin{eqnarray}
E_b^{(2)}=\frac{4\sigma}{( 2\pi \hbar)^2 \pi}
\int_0^{\pi}d\varphi \int_{0}^{\pi N_F} d \varphi_1 \nonumber \\
\varepsilon\Big( \varphi_1 +\varphi -\frac{\pi}{2}   \Big)
B\Big[\varphi; t(\varepsilon(\varphi_1 +\varphi-\pi/2))\Big]
 \label{EnergyAppendixB2B}
\end{eqnarray}

As the integration with respect to $\varphi$ is inside the interval
$(0, \pi)$ one may expand the area (as it was done in
Eqs.(\ref{S1},\ref{energyexpansion1})), and find
\begin{eqnarray}
\varepsilon( \varphi_1 +\varphi -\frac{\pi}{2}  )= \varepsilon( \varphi_1   ) +
\frac{2 \sigma}{m}(\varphi -\frac{\pi}{2} ).
\label{expansion}
\end{eqnarray}
Inserting this expansion in the above integral and performing integration with
respect to $\varphi$, after  simple but rather lengthy calculations one finds
\begin{eqnarray}
E_b^{(2)}=  \int_0^{\bar{\varepsilon}_F}  \varepsilon \nu(\varepsilon)
d \varepsilon     \hspace{4cm}        \nonumber \\
+ \frac{4\sigma}{( 2\pi \hbar)^2 }\Big\{ -\pi \bar{\varepsilon}_F \delta n_{F0} +
\frac{\hbar \omega_H}{2 \pi } (\pi \delta n_{F0})^2  \Big\}.
\label{EnergyAppendixE2final}
\end{eqnarray}
While writing the above equation we used the equality $\delta  \bar{n}_F=\delta n_{F0}$
(see Eq.(\ref{FEcouplingbis}) and the text below it).
%that follows from Eq.(\ref{conservlaw2}), as well as the definition of $\delta n$.

Inserting Eqs.(\ref{EnergyAppendixE1final},\ref{EnergyAppendixE2final}) in
Eq.(\ref{AppendixBandSum}) one obtains the band energy of the system with DW under
magnetic fields, Eq.(\ref{Eband}) of the main text.

\end{appendix}

\end{document}